\documentclass[prb,superscriptaddress,twocolumn,showpacs,preprintnumbers,%
amsmath,amssymb]{revtex4}

\usepackage{graphicx}
\usepackage{dcolumn}
\usepackage{bm}

\begin{document}
\title{ Domain-wall excitons and optical conductivity  in 
one-dimensional Wigner lattices}
\author{Matthias Mayr}
\altaffiliation[Current address:]{Department of Physics, University of Tennessee, 
Knoxville, Tennessee 37996, USA }
\affiliation{Max-Planck-Institut f\"ur Festk\"orperforschung,
             Heisenbergstrasse 1, D-70569 Stuttgart, Germany} 
\author{Peter Horsch}
\affiliation{Max-Planck-Institut f\"ur Festk\"orperforschung,
             Heisenbergstrasse 1, D-70569 Stuttgart, Germany}

\date{\today}

\begin{abstract}

Motivated by the
recent finding that doped edge-sharing Cu-O chain compounds 
like Na$_3$Cu$_2$O$_4$ and  Na$_8$Cu$_5$O$_{10}$  are
realizations of one-dimensional Wigner crystals, we study 
the optical spectra of such systems.
Charge excitations in  1D Wigner crystals are described in 
terms of domain wall excitations with fractional charge.
We investigate analytically and numerically the domain-wall excitations 
that dominate the optical absorption, and analyse the dispersion
and the parameter range of exciton states characteristic for the 
long-ranged Coulomb attraction between domain walls.
Here we focus on the Wigner lattice at quarter-filling 
relevant for Na$_3$Cu$_2$O$_4$ and 
analyze in particular  the role of second neighbor hopping $t_2$
which is important in edge-sharing chain compounds.
Large $t_2$ drives an instability of the Wigner lattice 
via a soft domain-wall exciton towards
a charge-density wave with a modulation period distinct from that 
of the Wigner lattice. 
Furthermore we calculate the  temperature dependence of the
DC-conductivity and show that it can be described by activated
behavior combined with a $T^{-\alpha}$ dependent mobility.

\end{abstract}
\maketitle

\section{\label{sec:introduction} Introduction }

At very low density an electron gas is expected to crystallize and
to form a Wigner crystal, as in this limit the Coulomb interaction 
among electrons dominates their kinetic energy \cite{Wigner34}.
In the  70's
Hubbard\cite{Hubbard78} as well as Kondo and Yamaji\cite{Kon77}
suggested, expanding the considerations by Wigner,
that the distribution of electrons in certain tetracyanoquinodimethane
(TCNQ) charge transfer salts may be controlled rather by the Coulomb interaction
than by the kinetic energy ($\sim$ band width), such that the electrons form
a {\it generalized} Wigner lattice (WL)  
on the underlying TCNQ chain structure.
This view suggests a strikingly different nature of charge excitations
namely domain walls with fractional charge rather than particle-hole 
excitations as in usual metals and semiconductors\cite{Hubbard78}.
This proposal, however, may be challenged on the grounds that
the resulting periodicity of charge modulation can alternatively be explained
by a $4k_F$ charge density wave (CDW)\cite{Eme76,Hir84,Sch93}
arising from  an instability of the
Fermi surface even in models with {\it short-range} interactions.
In fact there is only a gradual crossover 
between the WL and the $4k_F$ CDW, thus there is no clear distinction
between the two on the basis of the charge modulation period possible\cite{modulation}.

In a recent study of a new class of charge-ordered compounds
Na$_{1+x}$CuO$_2$ \cite{Sof05},
which contain
edge-sharing Cu-O chains, it has been suggested
that the magnetic and thermodynamic
properties of these compounds can only be explained in terms of
WL formation\cite{Hor05}.
Edge-sharing chains consist of CuO$_4$ squares just like  the Cu-O  planes of
high-T$_c$ cuprates, but they are differently linked. 
The edge-sharing arrangement of CuO$_4$ squares  meets the WL
criterion of small band width in an optimal way due to the almost
90$^o$ Cu-O-Cu bonds (Fig.1). Unexpected complexity is added
because, apart from a small nearest-neighbor hopping matrix
element $t_1$, the  second neighbor hopping $t_2$ has to be
considered which turns out larger as a consequence of the
structure. While this unusual feature does not affect the
classical WL order imposed by the Coulomb interaction, it changes
the Fermi surface topology\cite{kf}, and thereby allows to distinguish the
WL from the CDW on the basis of the modulation period.
The  Na$_{1+x}$CuO$_2$ compounds thus provide a first example where an unambiguous
distinction between the generalized WL and a Fermi surface related
$4 k_F$ CDW is possible. 

The electron interaction driven $4k_F$ CDW
has to be distinguished from the more familiar 2$k_F$ Peierls 
instability which arises from a modulation of hopping matrix elements
due to the coupling to periodic lattice distortions, i.e., leading to a CDW
centered on bonds rather than on the ions\cite{Gruener94}.
In contrast the Wigner lattice is based not 
on quantum but on classical energy considerations, namely which distribution
of localized electrons has the lowest Coulomb interaction. It is quite
remarkable though, that the periodicity of the WL coincides with that of
the 4$k_F$ CDW which emerges from a Fermi surface instability, i.e., 
from a pure quantum mechanical effect of strongly correlated electrons 
in models with nearest-neighbor hopping. 

Strictly speaking, at finite hopping $t_l$ ($l=1,2,\cdots$)
the WL is a {\it quantum solid}\cite{QS,Pol03}.
The kinetic energy causes virtual transitions to neighbor sites and 
leads thereby
to a quantum mechanical smearing of the electron positions.
Hence electrons should be rather visualized as electron clouds
that form a WL as consequence of the long-range Coulomb interaction.
This delocalization of the electron in the WL is a pure quantum effect, i.e.,
controlled by the interplay of kinetic and interaction energies.
The quantum mechanical bluring of electron position has
considerable effects on the physical properties of Wigner lattices,
as discussed in Ref.~\onlinecite{Hor05} in the context of edge-sharing chains in
 Na$_3$Cu$_2$O$_4$  and  Na$_8$Cu$_5$O$_{10}$ compounds,
where superexchange interactions and thereby the magnetic properties are strongly
influenced by the quantum nature of the WL.

The aim of the present paper is to explore the charge-excitations of the 1D WL,
which are characterized as domain-wall excitations.
We calculated the optical conductivity and its temperature dependence,
as it is hoped that forthcoming experiments may provide further evidence
for the WL nature of the electron structure of the 
Na$_{1+x}$CuO$_2$ edge-sharing chain compounds.
We use here both numerical, i.e., zero and finite-temperature diagonalization,
and analytical methods to arrive at a deeper understanding of the nature
of charge excitations in  1D Wigner lattices at quarter-filling.
In particular,
we find that the long-range (repulsive) Coulomb interaction among electrons
leads to exciton states below the domain-wall continuum, which appear as
strong exciton absorption in the optical conductivity in the case of
small hopping $t_1$.
Remarkably the second-neighbor hopping processes $t_2$ are found to
contribute strongly to the exciton dispersion.
These processes lift the degeneracy of the exciton state and the lower branch
gets soft at momentum $q=\pi/2$ and 
leads to an exciton instability at a critical value $t_{2,c}$.
The CDW state beyond  $t_{2,c}$ has a modulation period
twice as large as that of the WL. The charge modulation in the CDW
state is weak as compared to the WL state, 
as infered from  calculations of the static charge structure factor.

Doped edge-sharing chains are also building blocks of the intensively studied
system  Sr$_{14-x}$Ca$_x$Cu$_{24}$O$_{41}$, the so called telephone number 
compounds\cite{Mol89,Osa97,Blu02}. 
The composite structure of these materials consists of both ladder
and chain structures\cite{McC88}. While originally the attention was directed
toward the electronic properties of the ladders, because of their structural
similarity to the high-T$_c$ cuprates\cite{Dagotto96}, 
more recently the number
of papers reporting information concerning the chains is increasing.
The magnetic properties of these compounds which depend strongly on the doping 
are usually attributed to the chains\cite{Car96,Amm00,Kli05}.
Recently a quintupling of the chain unit cell in  Sr$_{14}$Cu$_{24}$O$_{41}$
due to charge ordering below $\sim 200$ K 
was reported\cite{Fuk02,Got03}.
An additional complexity of these compounds is due to the exchange of holes 
between chains and ladders\cite{Osa97}, 
i.e., the doping concentration of edge-sharing chains 
is difficult to infer precisely. Moreover an incommensurate modulation 
results from a misfit between the unit cells of ladders and chains.
Work by van Smaalen\cite{Sma03} and particularly a neutron scattering study  of
Braden {\it at al.}\cite{Bra04} illuminate the subtle aspects of the
interplay between modulations of chains and ladders. 
It has also been argued that the misfit between chains and ladders may
modify the charge ordering and hence the spin structure 
on the chains\cite{Gel04}.  
Particularly remarkable is a study by
Isobe {\it et al.}\cite{Iso00}, who succeeded in resolving the internal 
charge structure of the charge modulation of the compound 
[(Sr$_{0.029}$Ca$_{0.971})_2$Cu$_2$O$_3]_{54}$[Cu O$_2 ]_{77}$
with a chain unit cell containing  77 Cu sites.
Inspection shows that the charge pattern found in the structure analysis
compares favorably with that expected for a generalized Wigner lattice.
Recently also the modulation of the charge density in the ladders of  
Sr$_{14}$Cu$_{24}$O$_{41}$ has been reported by Abbamonte {\it et al.}\cite{Abb04}.

Wigner crystals are prime examples for strongly correlated states in the
sense that electrons do the utmost to avoid each other in real 
space\cite{Faz99}.
In general strong correlations (i.e., large on-site interaction $U$)
and the associated reduction of kinetic energy are favorable for charge
localization.
As a consequence
the long-range Coulomb interaction may become relevant in
strongly correlated systems, i.e., leading to charge ordered states 
and WL order in higher dimensions at particular fillings.
Examples are the manganites at quarter-filling which reveal checker-board 
charge order\cite{Goo55,Bal05}
and the layered molecular crystals of the BEDT-TTF type\cite{Mck01,Cal02,Dre03,Gre05}.
Also charge stripes in high-T$_c$ compounds at 1/8 doping\cite{Tra95} 
reflect the 
interplay of strong correlations and long-range Coulomb interactions.

The outline of the paper is as follows: In Section II we describe the
Hubbard-Wigner model for edge-sharing chain systems and 
 introduce the corresponding spinless fermion Hamiltonian. Furthermore we analyse the 
resulting domain-wall interactions for both Coulomb and truncated interactions.
In Section III we present diagonalization results for the charge-excitation spectrum 
and analyse the emerging excitonic states in the case of the model with
Coulomb interaction and nearest-neighbor hopping. 
Here we also provide an analytical derivation 
both for the continuum and the exciton states.
The temperature dependence of structure factor, kinetic energy, and
optical conductivity calculated by means of exact diagonalization (ED).
Section IV deals with the significant changes of the excitonic states
introduced by second neighbor hopping.
These are analysed with help of the analytical solution.
In particular we show here that the soft exciton states near $q=\pi/2$
are visible in the optical conductivity at elevated temperature as 
mid-gap absorption.
Finally the DC-conductivity of 1D Wigner lattices is discussed in
Section V, while 
our conclusions are summarized in Section VI. 

\section{\label{sec:section II}Model }

\subsection{\label{sec:section IIa}Wigner lattices in doped edge-sharing 
chain compounds }

\begin{figure}
\includegraphics[width=8cm]{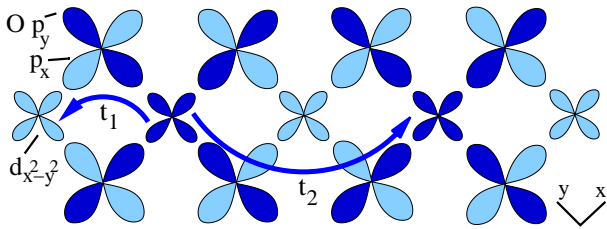}
\vspace*{4mm}
\caption{\label{structure1}
Orbital structure of edge-sharing copper-oxygen chains. 
The 90  degree Cu-O-Cu hopping $t_1$ and the second neighbor Cu-O-O-Cu
hopping path $t_2$ are marked by arrows.  
Shading indicates the p-d covalent mixing, as well as the hole distribution
in the $d^9L_h$ Zhang-Rice singlet states in  Na$_3$Cu$_2$O$_4$ 
with alternating charge order (dark shading). 
}
\end{figure}
As in the high-T$_c$ cuprates Cu$^{2+}$ is in a $d^9$ configuration
with spin 1/2, while Cu$^{3+}$ is in a $d^9$-ligand hole ($d^9 L_h$) singlet state,
i.e. involving the O neighbors of the Cu ion,
well known  as Zhang-Rice singlet\cite{Zhang88}.
Yet in contrast to the high-T$_c$ superconductors the edge-sharing geometry
with nearly 90$^o$  Cu-O-Cu bonds (Fig.\ref{structure1}) leads to strongly reduced 
hopping matrix elements. 
This sets the stage for the long-range Coulomb force as predominant interaction
\begin{equation}
H_{Coul}= U\sum_{i} n_{i,\uparrow} n_{i,\downarrow} 
+\sum_{i, l\geq 1} V_{l} n_i n_{i+l}  ,
\label{Hint}
\end{equation}
where the on-site interaction $U$ takes care of the strongly correlated character
of these systems and suppresses charge fluctuations leading to 
Cu$^{1+}$ ($d^{10}$) configurations.
In our model 
we associate the $d^9L_h$ ($d^9$,$d^{10}$) ionization state with
0 (1,2) electrons, respectively, and $n_{i,\sigma}$ ($\sigma=\uparrow,\downarrow$)
counts the number of electrons with spin $\sigma$, while 
$n_i=n_{i,\uparrow}+n_{i,\downarrow}$.
Thus the concentration of electrons $\rho$ is related to the concentration of holes 
$\delta=1-\rho$ (relative to the $d^9$ configuration) 
used in the high-T$_c$ literature \cite{rho}.
The Coulomb interaction $V_{l}$ is screened by the polarization of
neighbouring chains and by core electrons \cite{Hubbard78}. 
We shall not try to explore the subtleties of screening
due to the embedding, and assume for sake of simplicity a generic Coulomb
law  $V/l=\frac{V}{l}$, $l=1,2,....$, 
and keep the nearest neighbor interaction $V$ as a 
parameter\cite{periodic}.
Crucial for the following is that the interaction is long ranged and
convex, i.e., $V''_l=V_{l-1}-2 V_l+V_{l+1} > 0$.

\begin{figure}
\includegraphics[width=7cm]{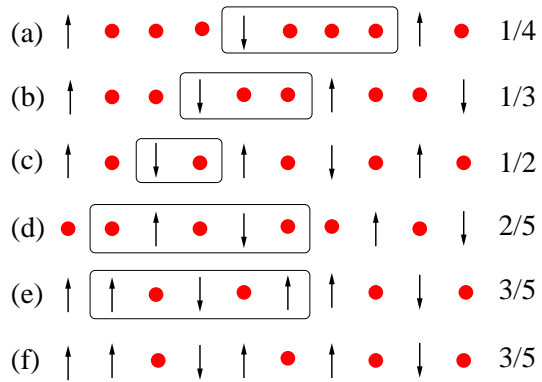}
\vspace*{4mm}
\caption{\label{wigner1}
Wigner charge order resulting from Coulomb repulsion
and associated modulated Heisenberg spin structure
for $\rho=1/4, 1/3, 1/2, 2/5$ and $3/5$ doping (a-e). 
The spin-1/2  of Cu$^{2+}$ (arrows) is responsible 
for magnetism, whereas Cu$^{3+}$ (red circle) is nonmagnetic.
The spin arrangement is that expected for ferromagnetic $J_1$
and antiferromagnetic $J_2$ exchange interaction. 
The charge unit cells (shaded) 
contain 4, 3, 2 and 5 sites, respectively.
The structures (c) and (d) are realized in   Na$_3$Cu$_2$O$_4$  and
Na$_8$Cu$_5$O$_{10}$, respectively.
A fluctuation of spin-position due to a low-energy charge excitation is
shown in (f) for $\rho=3/5$.
 }
\end{figure}

For commensurate doping concentration $\rho=m/n$ the interaction $V_l$
selects a particular charge ordering pattern \cite{Hubbard78}. 
The resulting charge order is
immediately obvious for the filling fractions $\rho=1/4, 1/3$ and 1/2 
(Fig.\ref{wigner1}(a-c)),
which involve an equidistant arrangement of the Cu$^{2+}$ sites
(arrows in Fig.\ref{wigner1}). 
For a general ratio $\rho=m/n$ this leads to complex structures with 
unit cell size $n$.
In case of $\rho=2/5$ and 3/5 we encounter in Fig.\ref{wigner1}(d,e)  
the charge order observed for   Sr$_{14-x}$Ca$_x$Cu$_{24}$O$_{41}$
and Na$_8$Cu$_5$O$_{10}$\cite{Hor05}.
Charge localization, however, is not perfect in Wigner insulators 
as electrons still undergo 
virtual transitions to neighboring sites (Fig.\ref{wigner1}(f))
in order to retain partially their {\it kinetic energy} . 
The energy of the lowest excitations and the impact of kinetic energy
depend strongly on $\rho=m/n$. For example, the energy
of the lowest excitation relative to the ground state 
Fig.\ref{wigner1}(c) is $\sim V''_2$
while the excitation for $\rho=3/5$ shown in Fig. 2(f) is 
$\sim V''_5$, about an order of magnitude smaller.
Hence quantum charge fluctuations are more important in the latter case\cite{Hor05}. 

To investigate the role of kinetic energy
we explore the dynamics of electrons starting from the 
one-dimensional {\it Hubbard-Wigner model} $H_{HW}=H_{Coul}+H_{Kin}$ \cite{Hubbard78},
where
\begin{equation}
H_{Kin}=-\sum_{i,l,\sigma} t_l(c^{\dagger}_{i+l,\sigma}c^{}_{i,\sigma}
+c^{\dagger}_{i,\sigma}c^{}_{i+l,\sigma})
\end{equation}
describes the hopping of an electron with spin $\sigma$ from site $i$ to site $i+l$
and vice versa; and $n_{i,\sigma}=c^{\dagger}_{i,\sigma}c^{}_{i,\sigma}$.  
Due to the almost 90 degree Cu-O-Cu angle the hopping  $t_1$ 
between nearest neighbor Cu sites results mainly 
from direct $d-d$ exchange, while $t_2$ originates from
hopping via a Cu-O-O-Cu path \cite{Mizuno98} (Fig. 1), leading to the remarkable
fact $|t_2|>|t_1|$.
We adopt here as typical values $t_1\sim 63$ meV, $t_2\sim 94$ meV,
derived from ab-initio  band structure calculations
for the Cu$^{2+}$ edge-sharing reference system Li$_2$CuO$_2$ \cite{weht98}.  
The hopping integrals
are much smaller than our estimates for $U\sim 3.8$ eV and $V\sim 1.5$ eV. 
Thus these parameters suggest that  the edge-sharing chains are well
inside the WL regime.

In particular  $|t_2|>|t_1|$ implies that the second-neighbor (antiferromagnetic)
exchange integral $J_2$ is large compared to the nearest-neighbor interaction
$J_1$ and the interchain couplings.
In fact this is consistent with the magnetic properties of the
 Na$_3$Cu$_2$O$_4$  and  Na$_8$Cu$_5$O$_{10}$ compounds\cite{Hor05} and also
with recent neutron diffraction data of the
compound NaCu$_2$O$_2$ which consists of $\rho=1$ edge-sharing chains\cite{Cap05}.
Magnetic excitation spectra determined by Raman spectroscopy \cite{Cho05} 
and a high-field NMR study\cite{Horvatic05} on  NaCu$_2$O$_2$
single crystals confirm these conclusions. 

It is evident that the spatial variation of charges
in the WL at the same time implies a complementary arrangement for the spins
and thus leads to {\it spatially modulated Heisenberg spin chains}
with varying distances among the spins\cite{Hor05,Sch05}.
This new category of spin models shows
very different magnetic properties for different
commensurabilities $\rho=m/n$, as actually observed in the 
Sr$_{14-x}$Ca$_x$Cu$_{24}$O$_{41}$ compounds\cite{Car96,Kli05}
and in the Na$_{1+x}$CuO$_2$ systems.
For example, Sr$_{14}$Cu$_{24}$O$_{41}$ has the commensurability
$\rho=2/5$ corresponding to the structure shown in Fig.\ref{wigner1}(d)
and its ground state is determined by singlet pairs\cite{Mat96,Fuk02,Got03,Kli05a}.
Thus its magnetic properties are quite distinct from those of 
Na$_3$Cu$_2$O$_4$ $(\rho=1/2)$ and  Na$_8$Cu$_5$O$_{10}$ ($\rho=3/5$)\cite{Hor05}.

\subsection{\label{sec:modelIIa} Spinless fermions and domain-wall interaction }

Our study of the charge structure and dynamics
will be based on the spinless fermion version of the
Hubbard-Wigner Hamiltonian $H=H_t+H_C$:
\begin{equation}
H=-\sum_{i,l\geq 1} t_l(c_i^{\dagger}c^{}_{i+l}+c_{i+l}^{\dagger}c^{}_{i})
+\sum_{i,l\geq 1}V_ln_in_{i+l}\ ,
\label{hamiltonian}
\end{equation}
The underlying assumption of spin-charge separation, i.e., the neglect
of the effects of spin-degrees of freedom on charge correlations
and excitations, can be rationalized
when starting from the 1D Hubbard model which is Bethe ansatz soluble
\cite{Lieb68}. That model has marginal spin-charge 
coupling\cite{Schulz98,Carmelo92,Stephan90,Gia04}
and its charge excitations are described by free spinless fermions\cite{Tohyama95}.
The addition of the long-range Coulomb interaction leads then to the
model of interacting spinless fermions\cite{Bhaseen04}.
We note, however, that in the 1D WL there are additional mechanisms
due to the long-range Coulomb interaction 
which influence the exchange interactions as discussed in Ref.~\onlinecite{Hor05}, 
and which induce some coupling between charge and spin.
Such effects will be neglected here.

\begin{figure}
\includegraphics[angle=0,width=6.5cm]{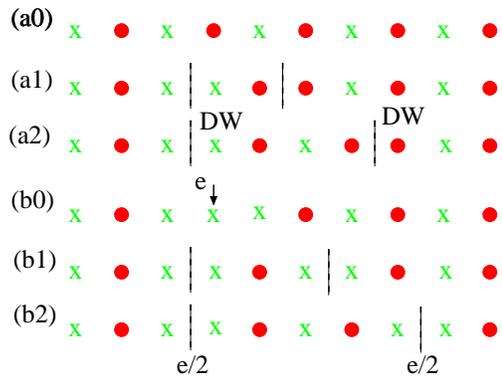}
\vspace*{4mm}
\caption{\label{fig:domain-walls}
(a) Charge excitations in a Wigner lattice move in form of domain walls 
(DW's). Here we consider the $\rho=0.5$ case
where Cu$^{2+}$ ( Cu$^{3+}$) are indicated by x(o), respectively.
(b) The addition of an extra electron 
with charge $e$ leads to a high energy state
which decays into two DW's with fractional charge $e/2$.
}
\end{figure}

In the following 
we shall focus on the $\rho=1/2$ case. The ground state has perfect
alternating charge order as shown in Fig.~\ref{fig:domain-walls}(a0) for $t_1=0$.
At finite hopping $t_1$ this ground state remains stable, yet
domain wall pairs get mixed in due to quantum fluctuations.
An elementary $t_1$ hopping process involves the interchange of 
(x0)$\rightarrow$(0x) pairs in  Fig.~\ref{fig:domain-walls}. 
The charge excitations in Wigner lattices, caused e.g. by optical excitations,
involve the creation of  domain wall (DW) pairs. 
These DW's can move as a consequence of $t_1$-processes of the kinetic energy. 
The role of $t_2$ processes is distinct in the $\rho=1/2$ case;
they are blocked in the perfect WL and contribute only in the
presence of DW pairs. We shall consider this problem in a later Section,
and assume for the moment $t_2=0$.


The charge of a DW can be invoked from a Gedanken experiment due to 
Hubbard\cite{Hubbard78}. The excitation generated by adding an 
extra electron with 
charge $e$ as shown in Fig.~\ref{fig:domain-walls}(b0)
will dissociate into two equivalent DW's with fractional charge $e/2$
separated by regions of perfect charge order. In general the domain
wall charge depends on the commensurability $\rho=m/n$.

The creation of a domain wall pair requires an energy $\propto V_1$.
Domain walls can propagate freely, yet due to the long-range interaction
$V_l$ they attract each other. For  $\rho=1/2$ the energy
of two domain walls $\Delta_m$ at distance $d=2m$ is determined by
the recursion relation
\begin{equation}
\Delta_{m} = \Delta_{m-1} + \sum_{n=m}^{\infty}V''_{2n}\;,\;\; m=2,3,\cdots
\end{equation}
with $\Delta_1 = \sum_{n=1}^{\infty} V''_{2n}$.
Here $V''_l=V_{l-1}-2 V_l+V_{l+1}$ denotes the discrete second derivative of 
the interaction. In the following we shall assume Coulomb interaction
$V_l=V/l$, and furtheron use $V=1$ as unit of energy. In this case
$\Delta_1=2 \ln{2}-1$.

An excellent asymptotic expansion for the domain-wall interaction (DWI)
has been derived by Fratini {\it et al.}\cite{Frat03} for the
Coulomb case:
\begin{equation}
\Delta_m\approx 1/2-1/8 m+1/(4 m)^3 -\cdots. 
\end{equation}
Remarkably this expression has an accuracy of two digits even at $m=1$.
It is interesting to note, that the leading interaction term $-1/8m$ 
in the asymptotic expansion can be interpreted in terms of 
an effective Coulomb interaction between the 
fractional charges of DW's $q_{1,2}=\pm1/2$. That is, the DW interaction is given 
as $q_1 q_2/d=-1/8m$, 
where $d=2m$ is the distance between the DW centers.
Hence the interaction between the domain walls $\Delta_m$ provides a manifestation 
of the fractional charge of the domain walls.

\begin{figure}
\includegraphics[width=7.5cm]{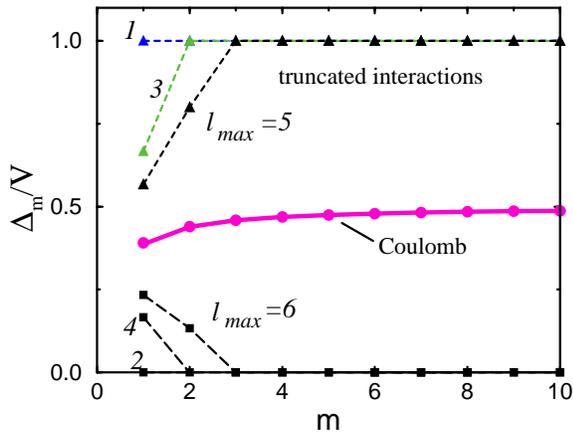}
\caption{\label{fig:DWint}
Interaction energy $\Delta_m$ of two DW's at distance $d=2m$ 
for a Wigner lattice with density $\rho=0.5$. Result for
Coulomb interaction (solid line) and asymptotic expression (circles)
is compared with
$\Delta_m$ obtained for Coulomb  interaction truncated at 
$l_{max}=1,2, .......,6$ (triangles and squares).
}
\end{figure}

It is evident that a truncated Coulomb interaction,
i.e. $V_l=0$ for $l>l_{max}$, may not stabilize the Wigner lattice structure
at general rational fillings \cite{Hubbard78,DeRaedt83,Hat97}.
The required value for $l_{max}$ increases with 
the commensurability ${\it n}$,  where $\rho={\it m/n}$ is the filling
fraction. In the case of the excitations truncation is even worse. 
This problem is demonstrated in Fig.~\ref{fig:DWint} 
which displays the energy $\Delta_m$
of two domain walls as function of distance $d=2 m$ for the most simple 
Wigner lattice $\rho=1/2$. In the case of Coulomb interaction $\Delta_m$
is an increasing function of $m$, i.e. the domain wall interaction is 
attractive, and converges against $\Delta_{\infty}=1/2$. Remarkably 
truncation leads to two different classes of behavior, repulsion at short
distance for $l_{max}=even$ and attraction for $l_{max}=odd$. Note that in the 
former case $\Delta_l\rightarrow 0$ for large $l$, while for $l_{max}=1$
which is sufficient to stabilize the Wigner lattice for $\rho=1/2$ there is
neither attraction nor repulsion, i.e., DW-pairs are not confined in this case.
Hence calculations of excitation spectra based on 
models with truncated interaction $V_1, V_2, ..., V_{l_{max}}$ 
or models with arbitraly chosen parameters $V_1, V_2$ etc. must be
considered with care when they are compared to the Coulomb case. 
A frequently studied model is the model with  both
nearest neighbor-hopping $t_1$ and nearest neighbor 
interaction $V_1$\cite{Hir84,Pen94,Cla03}.
The effects of longer-range interactions have been studied by Poilblanc
{\it et al.}\cite{Poilblanc97}, though for relatively weak interactions, i.e., 
outside the WL regime.

In this work we shall confine ourselves mainly to the discussion of 
the long-range Coulomb interaction. For the analytical considerations
we shall occationally consider the truncated  models with $l_{max}=1$
and $l_{max}=3$.

\section{\label{sec:section IV}RESULTS FOR NEAREST-NEIGHBOR HOPPING}

\subsection{Domain wall continuum and exciton}

\begin{figure}[t]
\includegraphics[width=8.0cm]{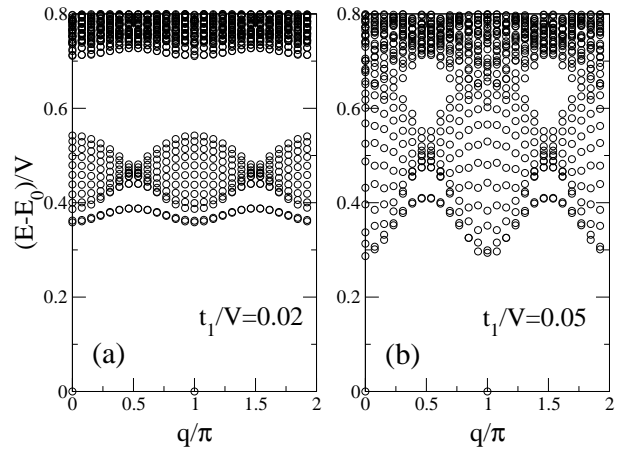}
\caption{\label{fig:excitations26}
Excitation spectrum for the quarter-filled chain with $N=26$ sites and Coulomb
interaction 
$(V=1)$ for (a) $t_1/V=0.02$ and 
(b)  $t_1/V=0.05$ as obtained by exact diagonalization.
The continuum due to domain-wall pairs is centered at $E\sim 0.45 V$ 
and its width is $\sim 8 t_1$. At small $t_1$ the
bound state is well separated from the bottom of the continuum, while for
larger $t_1$ (b) it merges with the continuum, but persists near $q=\pi/2$.
Parts of a continuum due to excitation of 4 domain walls is visible at the top
of the figure.
}
\end{figure}

As shown above
charge excitations in Wigner lattices 
form domain walls (DW's) separating
regions of perfect charge order.  DW's move as
a consequence of the kinetic energy operator $H_t$. 
For sake of transparency we consider first the model with
nearest-neighbor hopping motion $t_1$.
Numerical results for the excitation spectra for Coulomb interaction 
are given in Fig.~\ref{fig:excitations26}
for a quarter-filled ring (i.e. $\rho=1/2$) with $N=26$ sites. 
The figure shows the two degenerate ground-states at $q=0$ and $\pi$,
the 2 domain-wall continuum and part of the 4 DW continuum at high energy.
The 2 DW continuum is centered near $E\sim 0.45 V$
as expected from the DW interaction $\Delta_m\approx 1/2-1/8 m+1/(4 m)^3 -\cdots$
in the Coulomb case. 
The small downward shift from  $E=\Delta_{\infty}  =1/2$ is attributed 
to the interaction with the 4 DW states.
As a result of the Coulomb attraction between DW pairs an excitonic state
emerges below the 2 DW continuum, which is expected to play a prominent
role in the optical absorption.
With increasing hopping $t_1$ the width of the 2 DW spectrum $\sim 8 t_1$ increases 
and the exciton binding energy decreases, and 
eventually the exciton is absorbed
by the continuum near $q=0$ and $\pi$.
Yet due to the dispersion of both the lower edge of the 2 DW
continuum and the exciton, the excitonic state survives in the
vicinity of $q=\pi/2$.


\vspace*{0.0mm}

Next we  explore the analytical structure of the DW continuum
and then analyse the dispersion of the exciton in more detail.
We will address three different problems:
(i) The DW continuum in the case that all $\Delta_l$ are equal. This case
pertains to the model where only the n.n. interaction $V_1$ is kept.
There is no exciton state in this case and DW's are not confined\cite{Bhaseen04,Hat97}. 
(ii) Next we explore the solution for the bound state emerging for 
$\Delta_1<\Delta_{\infty}$
and $\Delta_l=\Delta_{\infty}$ for $l\geq 2$. This case directly applies to the model
where the Coulomb interaction is truncated at $l_{max}=3$, but this solution
also provides an approximate description for the case with full Coulomb 
interaction, if one adopts appropriate values for $\Delta_1$ and $\Delta_{\infty}$.
Finally (iii), we analyse the special role of the second neighbor hopping
$t_2$ in the case $\rho=1/2$.

We begin with the motion of a DW-pair as indicated in 
Fig.~\ref{fig:domain-walls}(a) and
denote the pair state by $|n,m>$. Here $n-m$ and $n+m$ denote the
centers of the xx and 00 DW's indicated by vertical bars in
Fig.~\ref{fig:domain-walls}, while $n$ denotes the center of mass coordinate
of a DW pair. 
We shall consider periodic boundary conditions, that is, 
even numbered rings of size $N$.
Then it is useful to introduce the auxilliary Bloch states
\begin{equation}
|\psi_{q,m}\rangle=\frac{1}{\sqrt{N}}\sum_n e^{iqn}|n,m\rangle;
\label{Blochbasis}
\end{equation}
where for periodic boundary conditions momenta are defined
as $q_{\nu}=2\pi \nu/N$ and $\nu=0,1, ......, N-1$.
When applying the translational operator $T_l$ one obtains
$T_l |\psi_{q,m}\rangle =e^{-iql} |\psi_{q,m}\rangle$.
The DW interaction energy associated with these states is $\Delta_m$ as discussed
above:
\begin{equation}
H_C |\psi_{q,m}\rangle=\Delta_m  |\psi_{q,m}\rangle.
\end{equation}
The action of the kinetic energy operator $H_t$ on the local DW states yields:
\begin{eqnarray}
H_t |n,m\rangle=
t_1\Bigl[\Bigl( |n-1&,&m+1\rangle \\ \nonumber
  +  |n+1&,&m+1\rangle\Bigr)(1-\delta_{m+1,N/2})\\ \nonumber
+\Bigl( |n+1&,&m-1\rangle\\ \nonumber
 +|n-1&,&m-1\rangle\Bigr)(1-\delta_{m-1,0})\Bigr].
\end{eqnarray}
This can be expressed in terms of the auxilliary Bloch basis 
Eq.(\ref{Blochbasis}) as:
\begin{eqnarray}
H_t |\psi_{q,m}\rangle= t_1(q) \Bigl[(1&-&\delta_{m+1,N/2}) 
|\psi_{q,m+1}\rangle \\ \nonumber
+(1&-&\delta_{m-1,0})|\psi_{q,m-1}\rangle\Bigr],
\end{eqnarray}
with $t_1(q)=2 t_1 \cos(q)$.
(i) In the case where all $\Delta_m$ have the same value, which we denote
$\Delta_{\infty}$, the solution for the domain wall continuum is straightforward:
\begin{equation}
|\Phi_{q,p}\rangle=\frac{2}{\sqrt{N}}\sum_{m=1}^{N/2-1} sin(pm)|\psi_{q,m}\rangle,
\end{equation}
where the $N/2-1$ pseudo momenta are determined by
$p_{\mu}=2\pi \mu /N$ and $\mu=1,2, ........, N/2-1$.
The corresponding energies of the DW continuum are: 
\begin{equation}
E_{q,p}= \Delta_{\infty}+2 t_1 \bigl[cos(q+p)+cos(q-p)]\bigr].
\end{equation}
That is, they are given as linear combinations of single domain wall energies.
The total width of the continuum is $8t_1$.

While the $t_1$-$V_1$ model with nearest-neighbor interaction stabilizes
the alternating charge-ordered ground state for $\rho=0.5$, it does not 
lead to an attractive interaction between domain walls. 
The excitation spectrum consists of the domain wall continuum. The
absence of the domain-wall exciton shows that domain walls are not confined
in this case\cite{Bhaseen04,Hat97}.


\vspace*{0.0mm}
\begin{figure}[h]
\includegraphics[width=8.0cm]{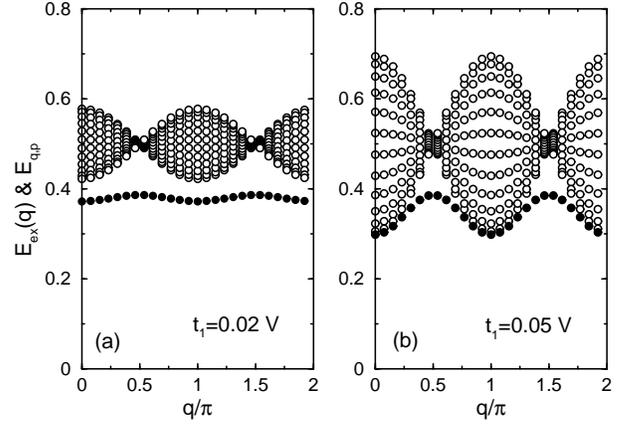}
\caption{\label{fig:excitations26_analytical}
Analytical solution for the domain wall continuum $E_{q,p}$
of the quarter-filled chain with $N=26$ sites and $\Delta_n=1/2$
for $t_1=0.02$ and 0.05. 
Exciton dispersion $E_{ex}(q)$
calculated for the model with $\Delta_1=2\ln{2}-1$ and $\Delta_m=1/2$
for $m\geq 2$ relevant for the case of long-range interaction.
}
\end{figure}

The simplest case which shows a bound state is the model where the
Coulomb interaction is truncated at $l_{max}=3$. 
The resulting DWI $\Delta_m$ in this case is $\Delta_1=\frac{2}{3} V$ and
$\Delta_m=\Delta_{\infty}=V$ for $m\geq 2$, i.e., the
the DW continuum is centered around $V$. 
Next we shall derive an analytical expression for the exciton energy.

The exciton solution can be obtained by expansion of the secular determinant
\begin{equation}
D=\left|
\begin{array}{cccccc}
\label{eq:D}
\omega_1 & t_1(q) & 0 & 0&\ldots& 0\\
t_1(q)&\omega_2   & t_1(q) & 0 &\ldots& 0\\
0 &t_1(q)&\omega_2   & t_1(q) &\ldots& 0\\
 \vdots &\vdots &\vdots &\ddots &\vdots &\vdots \\
 0 &\ldots& 0 & t_1(q) &\omega_2 &t_1(q)\\
 0 &\ldots & 0 & 0 & t_1(q) &\omega_1 \\
\end{array}
\right |
\end{equation}
where $\omega_l=\Delta_l-E$ and $t_1(q)=2 t_1 cos(q)$.
\begin{equation}
\label{eq:D_b}
D=\omega_1^2 D_M - 2\omega_1 t_1(q)^2 D_{M-1} + t_1(q)^4 D_{M-2}
\end{equation}
\begin{equation}
D_M=\left|
\begin{array}{cccc}
\omega_2 & t_1(q) & \ldots &\ldots\\
t_1(q)&\omega_2   & t_1(q) &\ldots\\
 \vdots &\vdots &\ddots &\vdots  \\
 \ldots&\ldots& t_1(q) &\omega_2 \\
\end{array}
\right |
\end{equation}
\begin{equation}
D_M=\omega_2 D_{M-1} - t_1(q)^2 D_{M-2},
\end{equation}
where $D_1=\omega_2$, $D_2=\omega_2^2-t_1(q)^2$ and $D_0=1$. 
The last relation is required so that the previous determinantal equation is fulfilled 
for $M\geq 2$. The equation may be solved by the polynomial ansatz
$D_M=p^M$, leading to
\begin{equation}
p=\frac{1}{2}\Bigl[\omega_2\pm\sqrt{\omega_2^2-4 t_1(q)^2}\Bigr]
\end{equation}
The equation for the bound state $\omega_1 p=t_1(q)^2$ takes then the form:
\begin{equation}
(\Delta_1-E)\Big[\Delta_2-E\pm\sqrt{(\Delta_2-E)^2-4t_1(q)^2}\Bigr]=4t_1(q)^2
\end{equation}
The physical solution, which is twofold degenerate because of Eq.(\ref{eq:D_b}),  
is
\begin{equation}
E_{ex}(q)=\Delta_1-\frac{t_1(q)^2}{\Delta_2-\Delta_1},
\label{Eex}
\end{equation}
where the dispersion $t(q)=2 t_1 \cos(q)$, which appears squared, reflects
the hopping of domain walls by two lattice units.
We also notice that the dispersion is weighted by the DW binding potential
$\Delta_2-\Delta_1$. As this quantity is $V/3$ in the model 
with truncated interaction ($l_{max}=3$), the dispersion of the exciton
(not shown) is smaller than in the model with Coulomb interaction.

It is instructive to compare the analytical results obtained so far 
with the numerical data obtained for Coulomb interaction  
displayed in Fig.~\ref{fig:excitations26}.
We approximate $\Delta_m$ as follows
$\Delta_1=2\ln{2}-1$ and $\Delta_m=1/2$ for $m\geq 2$.
The results are shown in Fig.~\ref{fig:excitations26_analytical}.
Although the approximation used for the DWI $\Delta_m$ is rather crude, 
there is quite a good agreement with exact diagonalization.

\subsection{\label{sec:melting of WL}Structure factor and melting of Wigner Lattice}

\begin{figure}
\includegraphics[angle=-0,width=8.5cm]{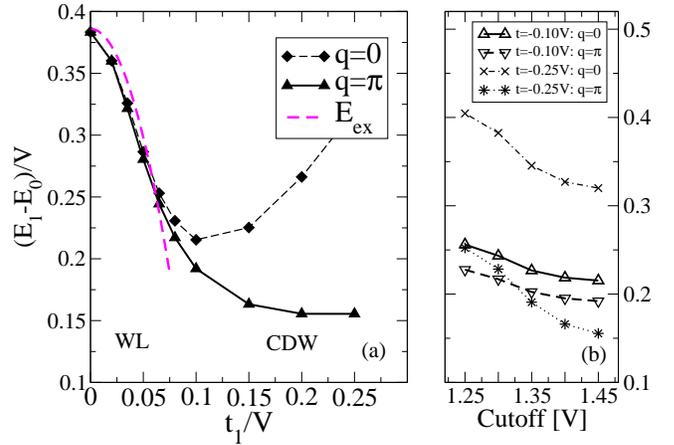}
\caption{\label{fig:charge-gap}
(a) The variation of the charge gap at $q=\pi$ as function of $t_1$
displays the crossover from the Wigner lattice to the CDW regime
in the case with nearest-neighbor hopping $t_1$.
Results are obtained for a $N=26$ site ring using  an energy cutoff
1.45 $V$. Exciton energy of WL (dashed curve) according to Eq.(\ref{Eex}).
(b) Dependence of the gap energies for large $t_1=0.1 (0.25)$ on the
cutoff energy.
}
\end{figure}

Wigner lattice order gets weakened with increasing 
kinetic energy $\sim t_1$. Yet in the model with only nearest neighbor hopping
there is a {\it gradual crossover} to the CDW  with weaker charge order
but the same modulation period\cite{Cap00}, which is $q=\pi$ at $\rho=1/2$.
This is the  $2k_F$ CDW of spinless fermions, and corresponds
to the $4 k_F$ CDW\cite{Hir84,Sch93} if spin is included in the model.
The crossover from the WL to the CDW regime is reflected in the variation 
of the charge gaps at $q=\pi (0)$ displayed in Fig.\ref{fig:charge-gap}(a)
as function of $t_1$.
While the two gaps are almost equal in the WL regime at small $t_1$,
for larger $t_1$ the $q=\pi (=2 k_F)$ charge gap controls the CDW state.
In this figure one can also see that in the WL regime the charge gaps
calculated numerically for a $N=26$ site cluster are well described
by the perturbative expression, Eq.(\ref{Eex}),
for the exciton energy $E_{ex}$  using $\Delta_1=2\ln{2}-1$ and $\Delta_2=1/2$
appropriate for the WL stabilized by Coulomb interaction.  
We note, that for the model with nearest neighbor interaction $V$ the CDW
solution for $\rho=1/2$ is expected to be stable up to $t_1=0.5 V$\cite{Cla03}, at even 
higher values for $t_1$ a metallic state will be realized\cite{Poilblanc97}.

Next we shall investigate the disappearence of WL order due to thermal 
fluctuations. It is well known that one-dimensional systems controlled by short-range
interactions do not exhibit long-range order at finite temperature.
The corresponding absence of a phase transition at finite temperature
follows from the Mermin-Wagner theorem\cite{Mer66,Bru01}. 
However, there is a {\it finite} transition temperature even in 1D models if the
decay of interactions is power-law like and sufficiently slow,
as has been shown for Heisenberg and Ising systems with ferromagnetic\cite{Dys69}
and antiferromagnetic\cite{Erwin05} interactions. The latter work 
by Erwin and Hellberg
which deals with an interaction decaying as $1/R$ is of direct relevance
for the WL case.

\begin{figure}
\includegraphics[angle=-90,width=8.0cm]{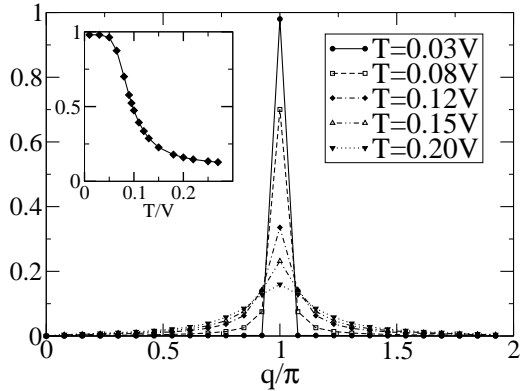}
\caption{\label{fig:Sq}
Structure factor $N(q)$ versus $q$
for a $N=26$ chain with Coulomb interaction  and $t_1=0.02 V$ for different 
temperatures. Inset shows the temperature dependence of $N(q=\pi)$.
}
\end{figure}

The melting of the generalized WL\cite{melting} 
due to domain wall excitations is reflected in
the structure factor $N(q)$, whose temperature dependence we study here.
The $T=0$ charge structure factor
\begin{equation} 
N(q)=(1/N_e^2)\sum_{i,j}\exp^{iq(i-j)}\langle n_i n_j\rangle 
\label{Nq}
\end{equation}
is normalized here
with respect to the total electron number of electrons $N_e$, such that
$N(q=\pi)=1$ for perfect CO, that is, at low temperature and for small $t_1$.
The alternating charge order is reflected in the peak of $N(q)$ 
at $q=\pi$ which starts to decrease strongly for $T>0.05 V$.
Yet even at much higher temperatures $N(q)$ reveals a maximum
at $q=\pi$ which indicates the persistence of pronounced short range 
charge correlations.
The results for $N(\pi)$ displayed in the inset
in Fig.~\ref{fig:Sq} indicate the melting of Wigner charge order\cite{Tm}
at a temperature $T_m \sim 0.06 V$ , i.e., for the case of Coulomb interaction
and $t_1=0.02 V$. 

We note that  $T_m \sim 0.06 V$ ($V\sim 1$ eV) has the correct order of magnitude
as the charge order transition $T_m\sim 455$ K in the compound 
Na$_3$Cu$_2$O$_4$  \cite{Hor05}.
Nevertheless we are far from any quantitative comparison, as 
3D effects could lead to a further enhancement of $T_m$ or to a reduction
due to frustration effects resulting, e.g., from the interaction with
Na-ion potentials.

\vspace*{00mm}

\vspace*{00mm}

Before we move on to the calculation of the optical conductivity,
we briefly comment on the exact diagonalization scheme used.
In the WL regime the eigenstates are energetically ordered according to
the number of domain-wall pairs that are excited 
(c.f. Fig.\ref{fig:excitations26}). 
This property of the Coulomb interaction, i.e., that it removes the 
high degeneracy of the usual Hubbard model,
allows to truncate the Hilbert space by selecting configurations with a small number
of domain walls, and to perform a full diagonalization in the
truncated space defined by a cutoff energy.
This procedure is expected to work well in the WL regime, but it should become
less accurate in the CDW regime. As the CDW can be visualized rather as
a Fermi sea with a small gap due to the relevant $2k_F$ scattering
processes, a truncated domain wall basis will certainly not be appropriate or 
require an extremly large cutoff energy such that again the full 
Hilbert space is covered.
Figure \ref{fig:charge-gap}(b)  shows the convergency of the lowest
excitation energies at $q=0$ and $\pi$ as function of the cutoff energy.
As can be seen from the figure, this truncation works extremely well
even for $t_1=0.1 V$, i.e., at the crossover to the CDW regime.
Thus most of our calculations for the bigger $N=26$ clusters employ
an energy cutoff $1.45$. Also a basis of momentum eigenstates is
used to further reduce the size of the matrices.


\subsection{\label{sec:optics}Optical conductivity }
\vspace*{0mm}
\begin{figure}
\includegraphics[width=7.8cm]{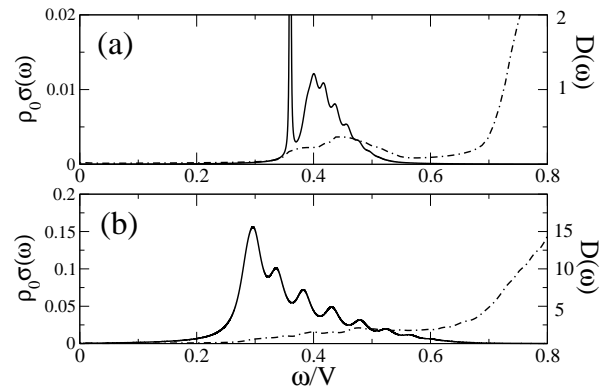}
\caption{\label{fig:optcond26_a}
Optical conductivity $\rho_0 \sigma(\omega)$ at $T=0$ (solid line) and 
density of states $D(\omega)$ of 2- and 4-DW excitations (dashed line) 
for  a $N=26$ chain with Coulomb interaction
for $t_1/V=0.02$ (top) and 0.05 (bottom), respectively. 
The spectra are shown with a 
Lorentzian broadening $\gamma=0.03$, while for the bound state in
$\sigma(\omega)$  $\gamma=10^{-4}$ was used. The strong increase of
$D(\omega)$ at $\omega\sim 0.7$ marks the onset of the 4-DW
continuum which does not contribute to $\sigma(\omega)$.
}
\end{figure}
Optical conductivity experiments could provide important information about
the size of the charge gap, the presence or absence of exciton features,
and the shape and width of the domain-wall continuum. Thus
in combination with the theoretical spectra 
optical data may allow to determine
the basic parameters of the Hubbard-Wigner model more precisely.
As we are dealing with insulating systems it suffices to focus on
the finite frequency absorption $\sigma(\omega)$ as given by
the Kubo equation in terms of  the current-current correlation 
function\cite{Mah90}
\begin{equation}
\sigma(\omega)=\frac{1-e^{-\omega/T}}{N\omega} Im\sum_{m,n}
\frac{e^{-E_m/T}|\langle n | j_x |m\rangle|^2}{\omega-(E_n-E_m)+i0^+},
\label{sigma}
\end{equation}
where $\langle n|$ ($\langle m|$) are eigenstates with energy $E_n$ ($E_m$), 
respectively. 
The current operator $j_x$ for the lattice model is defined in the usual way as
\begin{equation}
j_x = -\sum_{i,l\geq 1}\delta_l t_l(c_i^{\dagger}c^{}_{i+l}-c_{i+l}^{\dagger}c^{}_{i}), 
\end{equation} 
where $\delta_l=la$ is the hopping length, and $a$ the Cu-Cu distance along the
chain. 
The conductivity will be given in dimensionless form\cite{Mack99}, 
$\rho_0*\sigma(\omega)$, with 
$\rho_0=\hbar {\it v}/ e^2 a^2$ where ${\it v}=a b c$ is the cell volume per
Cu-site, and $\hbar/e^2=4.1 k\Omega$ the von
Klitzing constant. 

In this section we consider nearest-neighbor hopping amplitudes only. 
Equation(\ref{sigma}) can be evaluated in a straightforward way 
via exact diagonalization. 
Results for $\sigma(\omega)$ at $T=0$  are given in Fig. \ref{fig:optcond26_a} 
for two different hopping matrix elements $t_1$/$V$ = 0.02 and 0.05, which 
reveal the broadening of the domain-wall continuum $\sim 8 t_1$
and the disappearance of the exciton peak at larger $t_1$ in the continuum.
At $t_1=0.02 V$ the exciton at energy $ 0.38 V$ has large weight and the
shape of the DW continuum is asymmetric and peaked at the lower 
frequency edge. The strong asymmetry of the continuum persists at larger
$t_1$, i.e., when the exciton is no longer visible.

Figure~\ref{fig:optcond26_a} also provides a comparison of $\sigma(\omega)$
with the total density of states:
\begin{equation}
D(\omega)=(1/N)\sum_k\sum_n\delta(\omega-E_{k,n})
\end{equation}
The latter quantity shows apart of the two-domain wall continuum, 
which reaches up to $\omega\sim 0.6 V$,
also the much larger two particle-hole continuum involving four DW's. 
The latter, however, is not optically active.
Comparison of $D(\omega)$ and $\sigma(\omega)$ indicates a strong enhancement
of the optical matrix element towards the lower edge of the 2 DW continuum, i.e.,
leading to the asymmetric shape of  $\sigma(\omega)$.

\begin{figure}
\label{op_fig1}
\includegraphics[width=6.5cm]{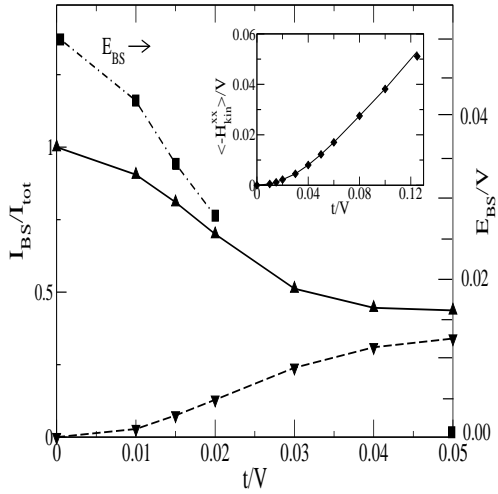}
\caption{\label{fig:weights}
Relative optical spectral weights of bound state (solid line) and of the first 
continuum state (dashed) versus $t_1$/$V$ calculated for a $N$=26 ring. The binding 
energy is given for small $t_1$ only because of the difficulty to extract this 
quantity from a discrete spectrum. The inset shows the kinetic energy (solid line) 
and the result obtained via the sum rule (triangles).  
}
\end{figure}

The weight of the lowest (exciton) excitation is compared in  Fig.~\ref{fig:weights} 
with the weight of the second lowest excitation. It can be
seen that the decay of exciton weight is rather fast with growing $t_1$ 
and parallel to the decrease of the exciton binding energy.
The inset of  Fig.~\ref{fig:weights} provides a comparison  of the kinetic
energy due to the quantum fluctuations and the integral over the
optical conductivity as function of $t_1$.
The coincidence of the data reflects the optical sum rule\cite{Bae86}:
\begin{equation}
\frac{2}{\pi}\int_0^{\infty} d\omega \sigma(\omega)
=\frac{1}{N}\sum_{l\geq 1}
\delta^2_{l}\langle -H^{l}_{kin}\rangle. 
\label{eq:sumrule}
\end{equation}
In the general case where different hopping processes in the kinetic energy 
$H^{l}_{kin},\; l=1,2, \cdots$, reach
over different distances $\delta_a$ the optical sum rule has to be modified 
and is given 
by an average over the individual kinetic energy contributions weighted
by the square of the hopping distances\cite{Aic02}. 
In all the cases considered thus far, the optical sum rule is fully exhausted by 
the incoherent part of $\sigma(\omega)$,  thus there is no 
zero-frequency Drude contribution.  
The increase of total spectral weight with $t_1$ naturallly follows from the
corresponding increase of kinetic energy. The quadratic dependence $\sim t_1^2/V$
at small $t_1$ highlights that the kinetic energy is due to virtual 
charge excitations with energy proportional to $ V$.

\vspace*{0mm}
\begin{figure}
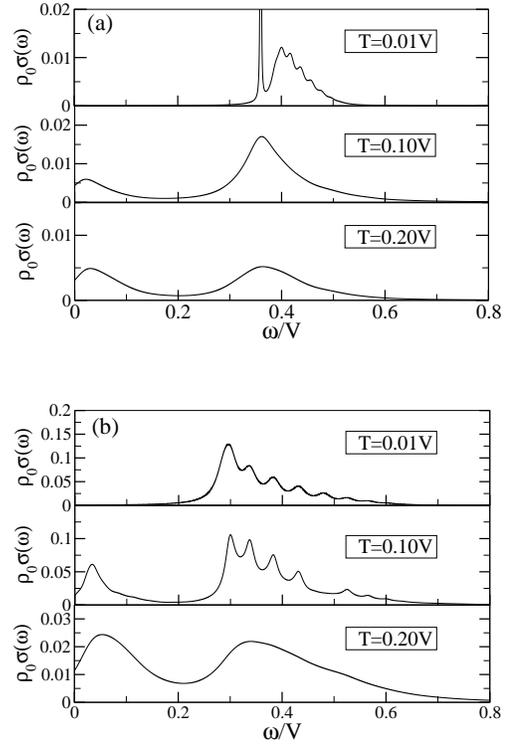

\label{op_fig2}
\includegraphics[width=6.5cm]{Fig_11a.eps}

\vspace*{8mm}
\includegraphics[width=6.5cm]{Fig_11b.eps}
\caption{\label{fig:optics_t05}
Optical conductivity at quarter-filling ($\rho=0.5$): 
(a) $t_1=0.02 V$ and (b)  $t_1=0.05 V$ for different temperatures
$T=0.01, 0.10 $ and $0.20 V$. 
}
\end{figure}

Next we consider the thermal evolution of the optical conductivity spectra
displayed in Fig.~\ref{fig:optics_t05} for the two cases $t_1/V=0.02$ and 0.05.
The spectra shown were obtained for a $N=26$ site cluster and periodic 
boundary conditions.
There are two immediately obvious features: (i) The disappearance of the 
prominent exciton absorption in the $t_1=0.02 V$ case at high temperature,
and (ii) the spectral weight transfer from high energy into a low energy gapless
absorption.
Furthermore one also realizes a gradual decrease of the total sum rule with increasing 
temperature.

The appearance of the low energy absorption arises from the thermal population
of  excited states, i.e., exciton and continuum states, and subsequent transitions 
within the continuum. The evolution of the low-energy continuum dictates the
temperature dependence of the DC conductivity. We shall come back to this point
in section VI.

%
\section{Role of second neighbor hopping }
%

Edge-sharing chain compounds have the peculiar property that the magnitude
of the second neighbor hopping matrix element $t_2$ is larger that the
nearest neighbor matrix element $t_1$.
This has the important consequence that the Fermi surface topology is
changed, i.e., instead of two Fermi points there can  now be 
four Fermi points depending on the hole or electron concentration.
Therefore we have here a qualitatively new situation compared to the
pure $t_1$ case, where the modulations of the CDW arising from the Fermi
surface instability and of the WL coincide.
At sufficiently large values for $t_2$, i.e., relative to 
the Coulomb interaction strength $V$, we expect the system 
to undergo a transition from the Wigner phase into a CDW state with different
modulation, i.e. now dictated not by the classical Coulomb interaction but
by `Fermi surface nesting', that is a charge modulation of
quantum mechanical origin. 

\begin{figure}
\includegraphics[width=6.5cm]{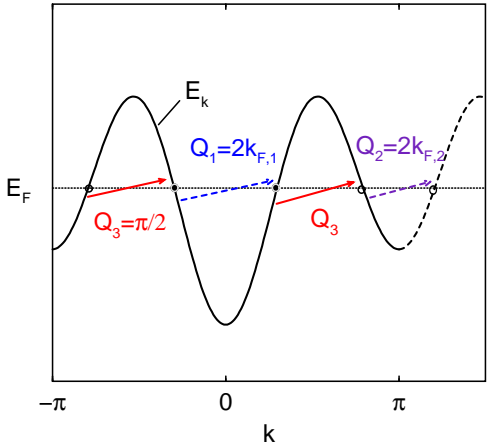}
\caption{\label{fig:Ek}
Free fermion dispersion $E_k$ for $t_2/t_1=2.5$ with scattering 
processes $Q_1=2 k_{F,1}$, $Q_2=2 k_{F,2}$ and $Q_3=\pi/2$ 
(for $\rho=1/2$) indicated by arrows.
}
\end{figure}

For sufficiently large $t_2$ there are
three relevant scattering processes $Q_1=2k_{F,1}$ and
 $Q_2=2k_{F,2}$ which are in general related to incommensurate
modulations determined by the ratio $t_2/t_1$, and a commensurate
modulation $Q_3=\pi/2$  as shown in Fig.~\ref{fig:Ek}.
Whereas $Q_1$ and $Q_2$ lead to the opening of gaps at two Fermi points,
respectively, $Q_3$ generates gaps at all 4 Fermi points simultaneously. 
Our numerical results show that the $Q_3$ modulation is the dominant one,
and leads to a further doubling of the unit cell.
It is remarkable that the modulation $Q_3$ coincides with that
of the $2k_F$-Peierls instability at quarter-filling, that is in the
usual model with nearest-neighbor hopping and spin.
Here we shall not follow the weak-coupling route further, but investigate 
the transition to the CDW from the strong-coupling WL side.

\vspace*{0mm}
 
\begin{figure}
\includegraphics[angle=0.0,width=8.5cm]{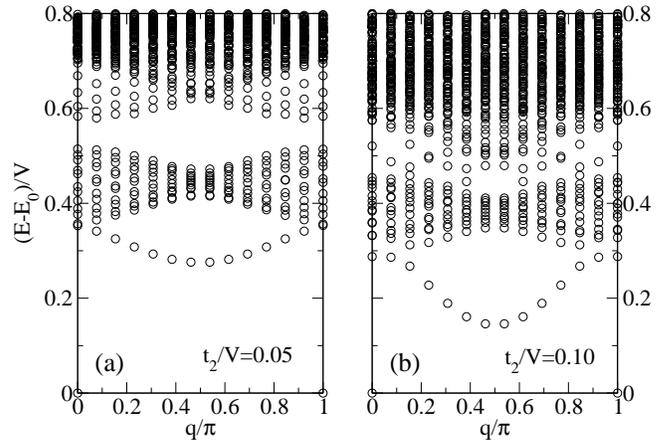}
\caption{\label{fig:exc_N26_t02_t2var}
Excitation spectrum for $N=26$, $t_1=0.02 V$ and two different second
neighbor hopping matrix elements $t_2=0.05$ and $0.10 V$,
showing the strong effect effect of $t_2$ on the exciton dispersion
with a shift of the minimum to $q=\pi/2$.
\vspace*{5mm}
}
\end{figure}

It is immediately evident that in the perfect charge ordered state
$t_2$ processes are completely quenched, and only through the presence
of charge fluctuations introduced via nearest neighbor hopping $t_1$ the
$t_2$ processes are activated.
This is seen in the excitation spectra for the Coulomb chain shown in
Figs.~\ref{fig:exc_N26_t02_t2var}. 
The exact diagonalization results show even for $t_2=0.10$  hardly any effect
on the 2-DW continuum. A small downward shift of the 2-DW spectrum
is attributed to a broadening of the 4DW continuum.

The exciton, however, is changed in a surprising way, it disperses downward,
in contrast to the $t_1$ case studied before. The numerical solution
reveals two further aspects: 
(i) the periodicity of the exciton dispersion indicates nearest neighbor
hopping of DW's, quite in contrast to the $t_1$ motion where DW's hop over
two sites; and
(ii) the exciton dispersion does not depend on the sign of $t_2$.

\begin{figure}
\centerline{\includegraphics[angle=-0.0,width=8.0cm]{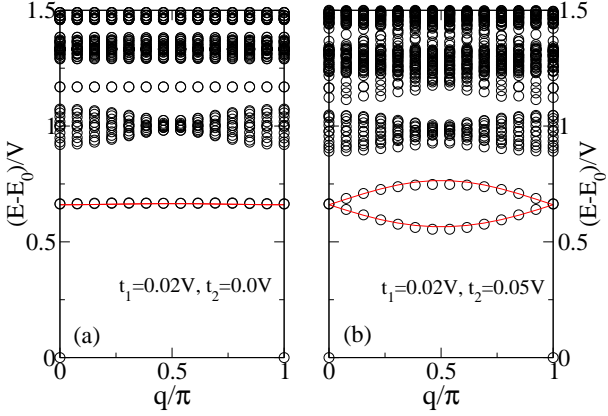}}
\caption{\label{fig:eq.V3}
Excitation spectrum for a $N=26$ site ring at electron 
concentration $\rho=0.5$
and Coulomb interaction truncated at $l_{max}=3$.
Numerical results (circles) for  
$t_1=0.02 V$ and two different second
neighbor hopping matrix elements: (a)  $t_2=0$ and (b) $0.05 V$.
Analytical result for domain-wall exciton states shown as solid lines.
}
\end{figure}

For further illustration we present in Fig.~\ref{fig:eq.V3} exact diagonalization
results for the excitation
spectrum of the model with an interaction truncated at $l_{max}=3$.
The domain wall interaction in this case is $\Delta_1=2/3 V$ and
$\Delta_m=\Delta_2=V$ for $m\geq 2$.
The DW continuum is centered around $V$. The dispersion of the exciton
in case (a) with $t_1=0.02 V$ and $t_2=0$ is strongly suppressed 
as compared to the case with long-range Coulomb interaction 
(cf. Fig.~\ref{fig:excitations26}).
As we shall see below, this is
due to the larger splitting $\Delta_2-\Delta_1$ between the exciton and the 
continuum as compared to the model with Coulomb interaction

At finite $t_2=0.05$ (Fig.~\ref{fig:eq.V3}(b)) the degeneracy of
the exciton is lifted and one observes clearly two exciton branches, one with
downward and one with upward dispersion.
To get some deeper insight in the peculiar dispersion of the exciton
at finite $t_2$, we turn now to the analytical analysis of the
excitonic state for the latter case, i.e., defined by the DW potential
$\Delta_1=2/3 V$ and $\Delta_m=V$ for $m\geq 2$.

\vspace*{0mm}

\subsection{Analytical study of the exciton state }

\begin{figure}
\centerline{
\includegraphics[angle=-0.0,width=4.5cm]{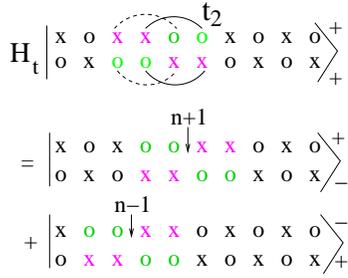}}
\caption{\label{fig:t2hopping}
Sketch of
2nd neighbor $t_2$ hopping processes between energetically degenerate and
symmetrized DW configurations. Such processes occur only in the $m=1$ sector
between symmetric and antisymmetric configurations (due to the Pauli principle)
and lead to a nearest
neighbor motion of the center of mass of the DW pair.
Signs attached to the right brackets indicate even and odd linear combinations.
}
\end{figure}

The analytical calculations presented above can be extended to the 
more realistic 
case of a finite next-nearest-neighbor hopping term $t_2$.
For $\rho=1/2$ $t_2$-hopping processes are completely 
blocked in the ground state with perfect charge order. A nearest-neighbor
DW pair, however, can move by one step to the left or to the right 
as a result of a $t_2$-process as indicated 
in Fig.~\ref{fig:t2hopping}. For the analytical solution it is useful to adopt
symmetrized configurations as shown in Fig.~\ref{fig:t2hopping}.

\vspace*{0mm}

There are only relevant matrix elements $\sim t_2$
between even and odd configurations in the $m=1$ sector
that are energetically degenerate:
\begin{equation}
_-\langle\psi_{q,1}|H_{t_2}|\psi_{q,1}\rangle_+ = t_2(q),
\end{equation}
where $t_2(q)=2 i t_2 sin(q)$.
This matrix element connects the even and odd sectors in the secular determinant:
\begin{equation}
D=\left|
\begin{array}{ccccccc}
\ddots & \ddots & 0 & \dots &\ldots& \ldots &0\\
t_1(q)&\omega_2   & t_1(q) & 0 &\ldots& \ldots & 0\\
0 &t_1(q)&\omega_1   & t_2(q) & 0& \ldots & 0\\
 0 &\ldots & t^*_2(q) &\omega_1 & t_1(q) & 0 & 0 \\
 0 &\ldots& 0 & t_1(q) &\omega_2 &t_1(q) & 0\\
 0 &\ldots & 0 & 0 & t_1(q) &\omega_2 & \ddots \\
 0 &\ldots & 0 & 0 & \ddots &\ddots & \ddots \\
\end{array}
\right |
\end{equation}
The expansion of the secular determinant yields a modified equation for the
bound states:
\begin{equation}
\Bigl[\omega_1 p -t_1(q)^2\Bigr]^2 - |t_2(q)|^2 p^2 =0
\end{equation}
The $t_2$-term leads to a splitting of the degenerate solutions 
obtained in the $t_1$-case into a lower and an upper branch:
\begin{eqnarray}
E_{ex,l}(q)&=& \Delta_1- |t_2(q)| -\frac{t_1(q)^2}{\Delta_2-\Delta_1 +|t_2(q)|},
\label{Eexl} \\
E_{ex,u}(q)&=& \Delta_1+ |t_2(q)| -\frac{t_1(q)^2}{\Delta_2-\Delta_1 -|t_2(q)|}.
\label{Eexu}
\end{eqnarray}
The coupling $t_2$ shifts the exciton minimum in the lower branch to $\pi/2$.
The exciton dispersion does not depend on the relative sign of $t_2$. 
At a threshold value $t_{2,cr}$ $q=\pi/2$
excitons condense and lead to the new state controlled by the kinetic energy.
The upper branch is a physical solution only as long as it does not touch
the DW continuum.

In  Fig.~\ref{fig:eq.V3} we provide a comparison of the analytical solutions
for the exciton dispersions, Eqs.(\ref{Eexl},\ref{Eexu} ), 
with exact diagonalization data 
for a $N=26$ site cluster in the case of Coulomb interaction truncated at 
$l_{max}=3$. As can be seen from Fig.~\ref{fig:DWint}  
in this case $\Delta_1=2/3 V$ and $\Delta_m=\Delta_2= V$ for $m\geq 2$.
The analytical exciton dispersions calculated from Eqs.(\ref{Eexl},\ref{Eexu} ) 
provide a good description of
the numerical values. The small deviations are mainly  due to a slight 
downward  shift of the two-DW continuum in the numerical calculation,
that results from its interaction
with the 4-DW continuum which is much broader in the $t_2=0.05$ case.
This also implies a small downward shift of the DW-exciton.

It is also instructive to compare Fig.~\ref{fig:eq.V3}(b) calculated for the
truncated interaction with  Fig.~\ref{fig:exc_N26_t02_t2var}(a) which was obtained
using the full Coulomb interaction.
In the latter case only the lower exciton branch can be seen and its dispersion
is slightly larger than in  Fig.~\ref{fig:eq.V3}(b) although the spectra
have been determined for the same hopping parameters.
These differences result from the different attraction of domain walls
$\Delta_2-\Delta_1$ in the two cases, which enter in the 3rd term on the
r.h.s. of exciton dispersion in Eq.(\ref{Eexl} ).

\subsection{Structure factor and exciton instability }

\begin{figure}[h]
\includegraphics[angle=-90.0,width=8.5cm]{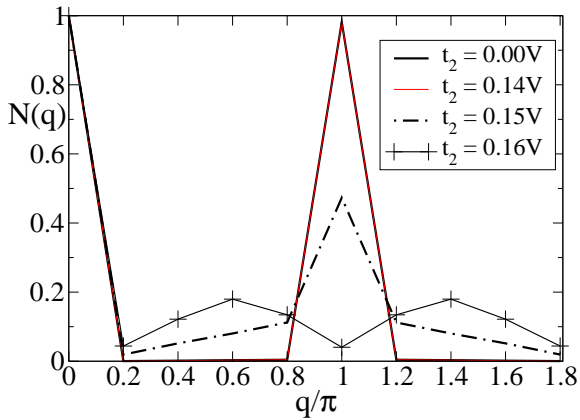}
\caption{\label{fig:Nq_N10_t2} 
Charge structure factor $N(q)$ as obtained by ED for a 10 site 
ring with Coulomb interaction ($V=1$) and $\rho=0.5$ 
for $t_1=0.02$ and  different values for $t_2$ ($T=0.01 V$). 
Straight lines are guides to the eyes;
the + symbols in the data for $t_2=0.16$  also indicate the
allowed $q$ points.
Wigner order is unstable beyond the critical value $t_{2}^{cr}=0.155$.
\vspace*{0mm}
}
\end{figure}

In the previous discussion of the electronic structure we have seen that
the exciton state will get soft at about $t_2=0.15 V$ and hence the
WL state should get unstable and a new ground state with momentum $q \sim \pi/2$
should appear.
In the following we analyse the change of the static charge structure factor.
Figure \ref{fig:Nq_N10_t2} 
shows the dependence of the charge structure factor $N(q)$, Eq.(\ref{Nq}),
on the size of $t_2$ for fixed $t_1=0.02 V$. 
The results are obtained for a $N=10$ site cluster where we included the full
Hilbert space in order to show the complete break-down of
Wigner order at large $t_2$.
\vspace*{0mm}

It is remarkable that up to the value $t_2\le 0.14$
the ground state correlations remain unchanged with $N(q)$ peaked at momentum $\pi$.
This reflects the blocking of $t_2$ hopping processes in the state with
alternating charge order.
Then in the narrow range $0.15>t_2>0.16$ there is a sudden change of $N(q)$
indicating a level crossing.
Beyond $t_2\sim  0.16$ the charge correlations are determined 
by the  new state with the maximum of
$N(q)$ near $\pi/2$.  The broad shape of $N(q)$ is  reminiscent of that
of a 1D Fermi gas, i.e., indicating that the CDW modulations in this phase
are rather weak. 
It has been checked by a calculation of the charge correlation functions
in real space, that there are still significant charge modulations in the
system consistent with a $q=\pi/2$ CDW.


\begin{figure}[h]
\includegraphics[angle=-0.0,width=8.5cm]{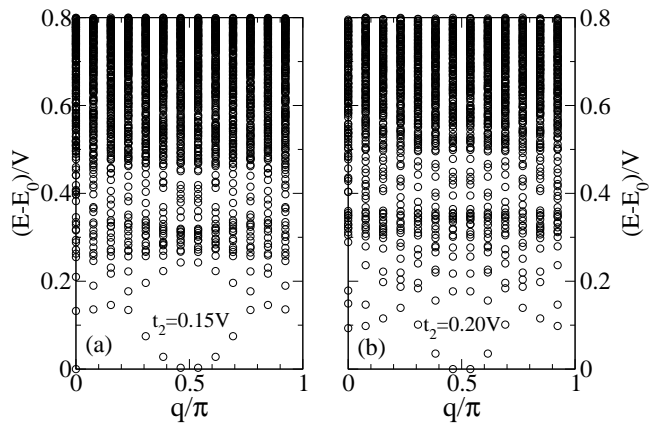}
\caption{\label{fig:eq_Vct1t2_002015}
Excitation spectrum for $L=26$, $t_1=0.02 V$ and two different second
neighbor hopping matrix elements $t_2=0.15$ and $0.20$ V indicating 
the exciton instability of the Wigner lattice.
While for $t_2=0.15$ the degenerate ground states are at $q=0,\pi$,
the ground state momenta for $t_2=0.2$ are near $q=\pi/2$.
}
\end{figure}

The excitation spectrum corresponding to the transition from the WL to 
the $\pi/2$ CDW is shown in Fig. \ref{fig:eq_Vct1t2_002015}(a).
It clearly reveals the quasi-degeneracy of the WL ground states
at $q=0(\pi)$ and the soft exciton at $q=\pi/2$.
At larger values for $t_2$ the state emerging from the exciton is the new
ground state of the system, as seen in  Fig. \ref{fig:eq_Vct1t2_002015}(b).
We note, that our program was developed for $N=4n+2$ site rings ($n=1,2, \cdots$)
that do not have $\pi/2$ as allowed momentum, therefore there are 
two degenerate ground states at $q=\pi/2\pm2\pi/N$.

The critical value $t_{2,cr}$ can also be estimated from Eq.(\ref{Eexl})
by setting  $E_{ex,l}(q)=0$ at $q=\pi/2$ and using the parameters
for the DW interaction appropriate for the model with Coulomb interaction.

\subsection{Optical conductivity }

\vspace*{0mm}
\begin{figure}
\centerline{\includegraphics[angle=-90.0,width=7.7cm]{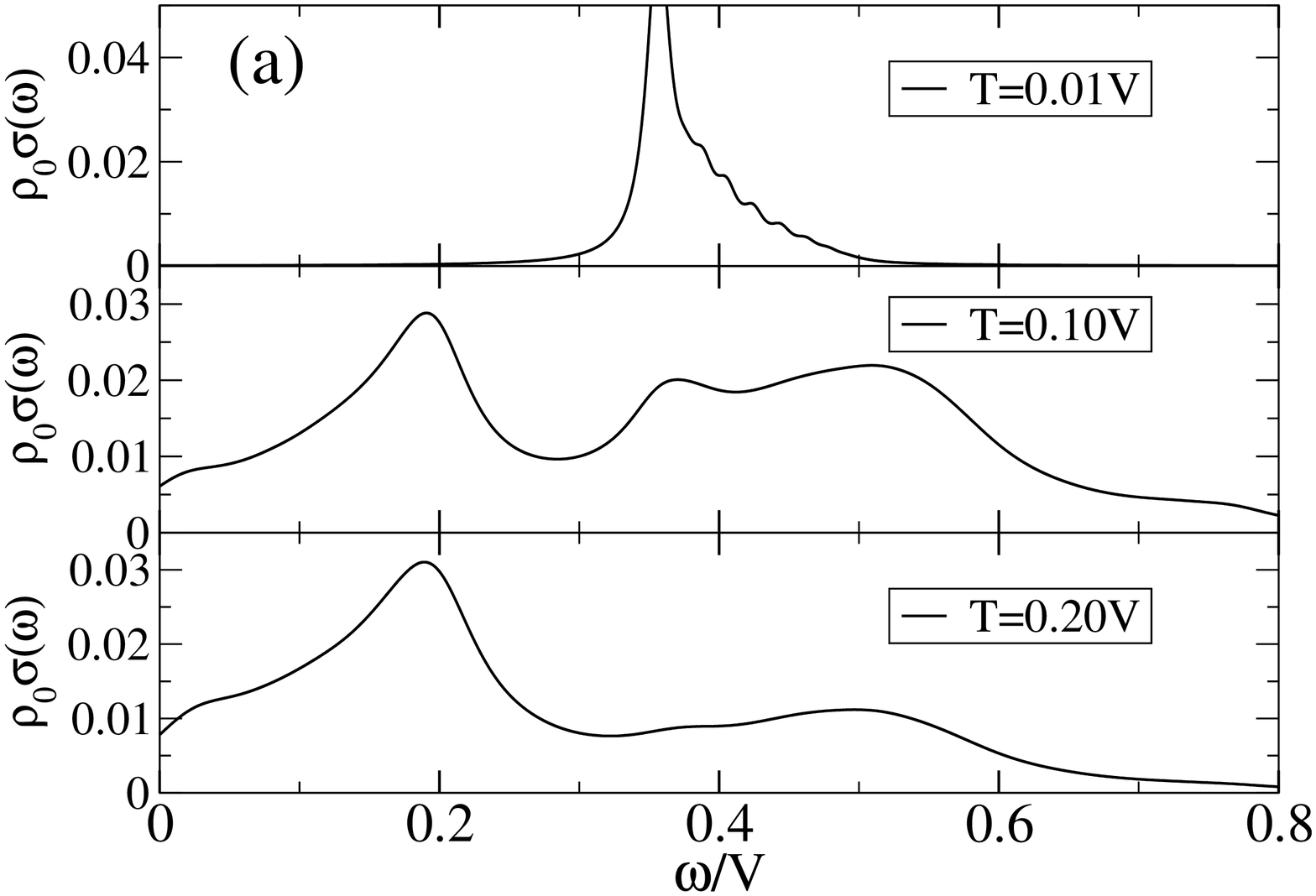}}
\centerline{\includegraphics[angle=-90.0,width=7.0cm]{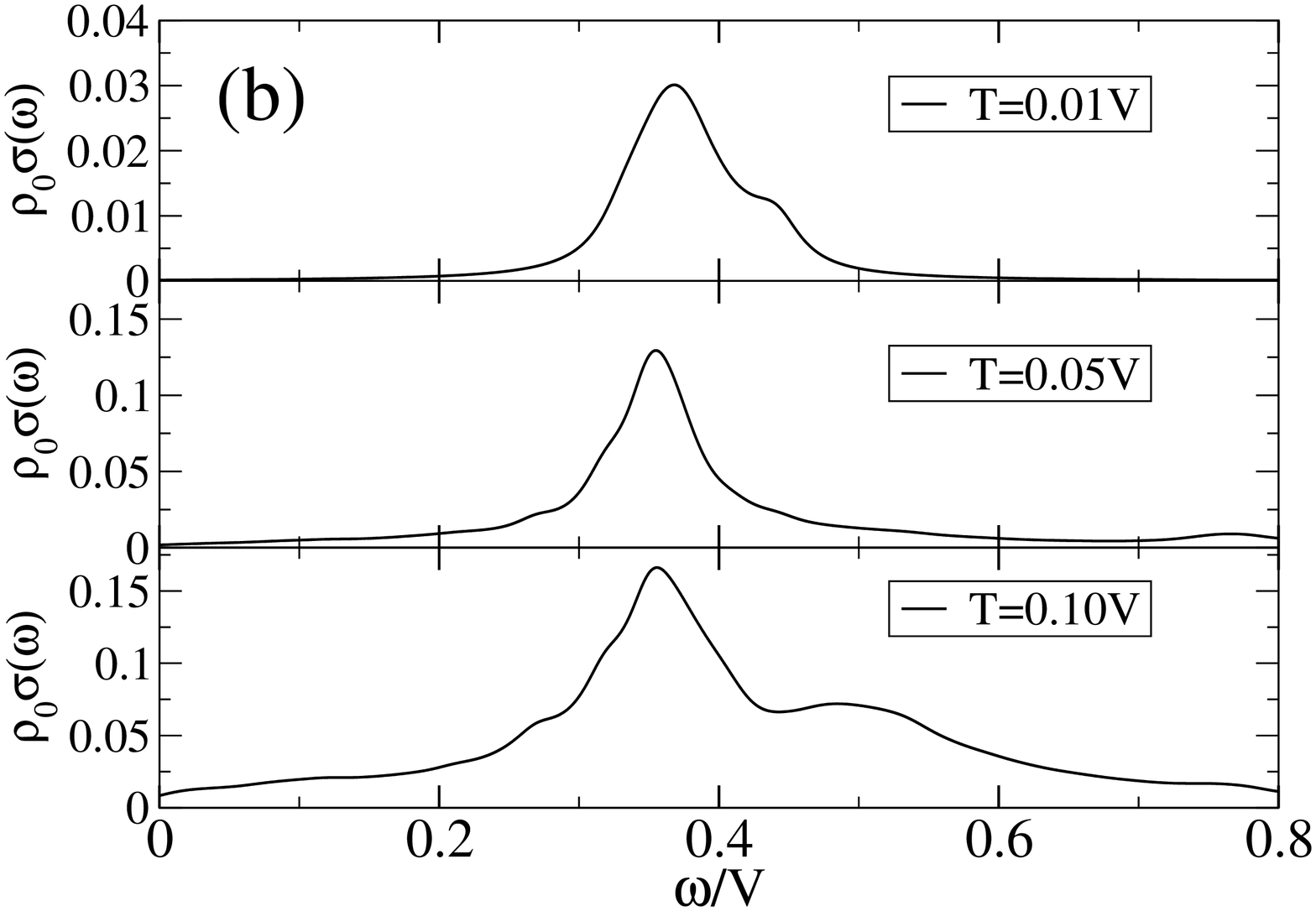}}
\caption{\label{fig:optcondN26}
Optical conductivity at quarter-filling ($\rho=0.5$) for different temperatures:
(a) $t_1=0.02, t_2=0.05$ (Lorentzian broadening $\gamma=0.05 V$), and 
(b) $t_1=0.02, t_2=0.10$ .
The in-gap ansorption appearing 
near $\omega\sim 0.19 V$  at finite temperatures in (a) is 
attributed to a transition from the thermally excited exciton 
in the vicinity of $q=\pi$
to the domain wall continuum. A similar feature appears in (b) at
$\omega\sim 0.27 V$.
}
\end{figure}

The effect of  second neighbor hopping 
$t_2$ on the optical conductivity $\sigma(\omega)$
is displayed in Figs.~\ref{fig:optcondN26}(a,b)
 for the model with Coulomb interaction. 
As a consequence of the blocking of $t_2$ processes in the state with
alternating Wigner charge order at $\rho=1/2$ the optical spectra
are only slightly modified at low temperature, and almost coincide
with those obtained for the model with only nearest-neighbor hopping 
Fig.~\ref{fig:optics_t05}. 
A remarkable change, however, is the disappearance of the $q=0$
exciton, which was very pronounced in the spectrum for $t_1=0.02$
and $t_2=0$ in Fig.~\ref{fig:optics_t05}(a) at low temperature.
This feature has disappeared
after switching on $t_2=0.05$ in  Fig.~\ref{fig:optcondN26}(a). 
For these parameters there is no bound state at $q=0(\pi)$ as can also be seen 
from the corresponding energy level diagram in Fig.~\ref{fig:exc_N26_t02_t2var}(a). 

As  an  optical experiment involves only
vertical transitions only charge excitations at $q=0$ and $\pi$ are
probed  at low temperature, i.e., the downward dispersing exciton is invisible. 
This changes, however, when the exciton states get populated by thermal excitations.
As a consequence
there are  marked changes in the spectra at higher temperature, which can be
traced back to the different dispersion of the exciton state at finite $t_2$.
The pronounced structure near $\omega=0.2$ in Fig.~\ref{fig:optcondN26}(a)
at $T=0.1 V$, which is already seen at $T=0.05 V$ as a weak 
{\it in-gap excitation}, stems from transitions between thermally
excited excitons near $q=\pi/2$ and final states at the upper edge of the 
2DW continuum  ( cf. Fig.~\ref{fig:exc_N26_t02_t2var}(a)).
This conclusion is based on a careful study of the size of the 
corresponding matrix elements in the current-current correlation function.
In the spectra of Fig.~\ref{fig:optcondN26}(b) for a twice as large
$t_2$ value this absorption has shifted
to higher energies ($\omega\sim 0.27 V$) and appears now as a small structure 
on the shoulder of the main peak, which is due to transitions into the
2DW continuum at $q=0$.

Thus the observation of an in-gap absorption in the optical conductivity
at finite temperature can provide valuable information about 
the position of the minimum of the
exciton dispersion in the middle of the Brillouin zone, which may 
allow to determine the value of $t_2$ from experiment.

Finally we recall that the higher transitions  into the upper Hubbard band
with energy $\sim U$ are not contained in the spinless fermion model we
study here. Yet they are contained in the Hubbard-Wigner model.
For $t_2\geq t_1$ and $\rho=1/2$ these transitons are expected at energy $U-V_2$ with
intensity $\sim t_2^2$. But also in the case $t_2 \ll t_1$ transitions into
the upper Hubbard band  
are expected due to charge fluctuations resulting from $t_1$ processes.


\section{DC-CONDUCTIVITY}
A central experimental quantity to compare our finite temperature results with 
is certainly the DC-conductivity $\sigma(T)$.
In Fig.~\ref{fig:optics_t05},
which displays the temperature dependence of $\sigma(\omega)$,
we observe the emergence of a low-frequency continuum whose intensity grows
with increasing temperature. These changes are accompanied by a spectral weight 
transfer from high to low energy.
The low-energy excitations arise from transitions within the 2 DW
continuum.
We note that at higher temperatures, i.e., near the melting temperature of 
the WL and above, also 4 DW excitations do contribute substantially to
the low-frequency absorption. This yields a very dense low-energy spectrum
even for small systems.
It is certainly suggestive that this low-frequency continuum contains
the information about the DC-conductivity.
Yet as we are dealing with finite systems the case is not that simple
as the zero frequency limit cannot be determined in a straightforward way.
In fact the analysis of the low-frequency part of   $\sigma(\omega)$ reveals
a pseudogap $\sim 8t/N$ which scales inversely with the size of the system $N$.

Similar pseudogap behavior and finite size effects in  $\sigma(\omega)$
have also been observed recently by Prelovsek {\it et al.}\cite{Prelovsek04} 
in a study of the 1D t-V model. 
In their careful study  Prelovsek {\it et al.}
arrived at the conclusion that after finite size scaling 
the results for frequency dependence could be compatible with a normal and 
featureless shape of $\sigma(\omega)$ as found by a frequency moment analysis.

\vspace*{10mm}
\begin{figure}[h]
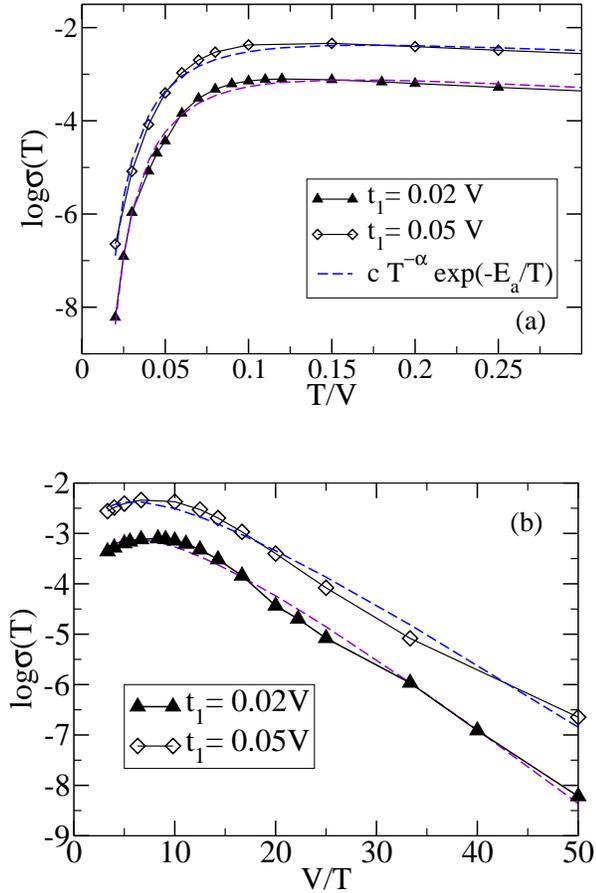

\includegraphics[width=7.5cm]{Fig_19a.eps}

\vspace*{5mm}
\includegraphics[width=7.8cm]{Fig_19b.eps}
\caption{\label{fig:sigma0vsT}
(a) Temperature dependence of the DC-conductivity for $\rho=0.5$ determined from the 
integrated low-frequency spectral weight for $t_1=0.02$ and 0.05 ($t_2=0$). 
(b)  Arrhenius plot  revealing the activated behavior at low temperature.
An activated behavior for the carrier density in combination with
a temperature dependent mobility 
$\sim T^{-\alpha}$, Eq.(\ref{fit}), provides a rather good fit
of the numerical data over the whole temperature range.
}
\label{sigma_DC}
\end{figure}

Thus we proceed as follows: 
We assume that in the thermodynamic
limit the level spacing vanishes and and the low frequency part of the
spectrum can be expressed by a Drude form for the real-part of the conductivity
\begin{equation}
\sigma(\omega)=\frac{\sigma(T)}{1+(\omega\tau)^2},
\label{Drude}
\end{equation}
where $\tau\approx 1/t_1$ is determined by the energy scale of the 
domain-wall continuum. The spectral weight of  the low-frequency part of 
   $\sigma(\omega)$ as determined by exact diagonalization
\begin{equation}
I(\omega_0)=\int_0^{\omega_0} \sigma(\omega) d{\omega},
\label{integral}
\end{equation}
with $\omega_0=4 t_1$ is then used to determine 
the DC-conductivity $\sigma(T)$ with help of Eq.(\ref{Drude}).

Results for the DC-conductivity obtained
from a 26-site ring at doping $\rho=0.5$ are presented in  
Fig.\ref{sigma_DC} for two different hopping matrixelements.
The numerical data for $t_1=0.02$ and $0.05 V$ ($t_2=0$) 
show similar behavior.
The DC conductivity reveals activated behavior below and basically
$T$-independent behavior above the melting temperature $T_m$ of the WL.
The conductivity for $t_1=0.02 V$ is smaller,  reflecting the $t_1^2$
proportionality of the current-current correlation function.

At small hopping $t_1=0.02 V$ the numerical data for $\sigma(T)$ is well 
described over 5 decades by a single activation energy dependence
$\sigma(T)\simeq c \exp{-E_a/kT},$
with $E_a\simeq 0.30 V$. However,
the sometimes used relation\cite{Epstein77,McCall87} 
\begin{equation}
\sigma(T)\simeq c T^{-\alpha} \exp{(-E_a/kT)},
\label{fit}
\end{equation}
which involves in addition
the temperature dependence of the mobility $\mu(T)\sim T^{-\alpha}$,
improves the fit when approaching the saturation (melting) regime at
high temperature as shown in Fig.\ref{sigma_DC} by dashed lines. 
Using this relation  we obtained for $t_1=0.02 (0.05) V$  the activation energy 
$E_a\sim 0.39 (0.32) V$, respectively, and $\alpha\sim 2.4 (1.95)$.
Thus the mobility $\mu(T)$ of the carriers (domain walls) decreases strongly
with increasing temperature, as one may have expected.
The activation energies $E_a$ determined from $\sigma(T)$ are consistent 
with the energy gaps in the corresponding energy level diagrams
in Figs.\ref{fig:excitations26}(a,b).
Yet the activation energies $E_a$ determined from $\sigma(T)$ are larger
than expected from the relation $E_a=E_g/2$  which applies for usual semiconductors. 
Whether this
discrepancy originates from the fact that the charge carriers
in the WL are domain walls with fractional charge and not usual
electron- and hole-like quasiparticles remains unclear and deserves further
study.

While the relation Eq.(\ref{fit}) provides a quite satisfactory description 
of the numerical data over the full temperature range, it is nevertheless far from
perfect. As one can see, the fit curve in Fig.\ref{sigma_DC}(b)
lies below (above) the numerical data for temperatures above (below) the
melting temperature $T_m$, respectively. This deviation possibly reflects the strong change
of the structure factor near $T_m$ (see inset of  Fig.\ref{fig:Sq}).
Here our aim was to keep the number of parameters as small as possible and therefore 
we have not tried to add such a $E_a(T)$ correction term that would involve $T_m$
and further parameters. 
 
Recent measurements of the DC-conductivity
of  Na$_3$Cu$_2$O$_4$ and  Na$_8$Cu$_5$O$_{10}$ compounds show the same trends: 
(i) an Arrhenius
behaviour below , and (ii) a saturation of DC conductivity above
the melting transition $T_m\sim 455 (540)$K, respectively\cite{Sofin03,Hor05}.
Moreover, the experimental conductivities show only a small discontinuity  
at the melting temperature $T_m$.
Although the absence of a discontinuity in the theoretical 
curve can be attributed to the finite system,  which is not
expected to display a phase transition, it is nevertheless remarkable
that also the experimental data does only show a weak discontinuity
at the melting temperature $T_{m}$ of the WL.
The measured DC-conductivity of the compound Sr$_{14}$Cu$_{24}$O$_{41}$
was also found to be described by an Arrhenius law with an activation energy 
$E_a\sim 0.12$ eV. More recent studies found a crossover between
two exponential regimes, with $E_a\sim 0.12$ eV for the low $T$ regime
up to about 170 K but different values 0.18  eV\cite{Blu02}, 0.22 eV\cite{Maeda03}  
and 0.27 eV\cite{Gor02} for the high-$T$ regime. 
Up to 400 K no saturation was observed for this compound. 
As the transport at high temperatures is probably  due to the
chains and the ladders in these compounds,
a direct comparison with our results is ruled out.


\section{CONCLUSIONS}

In summary, we have investigated the charge excitations of a 1D generalized
Wigner lattice, expected to be realized in edge-sharing Cu-O chain systems
and also in some organic chain compounds, starting from the
Hubbard-Wigner model with long-range Coulomb interactions $V/l$ among 
electrons.
A central aim was to gain insight into the spectral structure of the
optical conductivity $\sigma(\omega)$ and its temperature dependence.

We have found that:
(i)  $\sigma(\omega)$ is determined by a highly asymmetric spectrum due
to two domain-wall excitations with an energy gap $E_g\sim V/2 - 4 t_1$ and
a width of $\sim 8 t_1$. The asymmetric form of this spectrum
may serve as another fingerprint
to detect the WL and may also be employed to determine the parameters of
the model from experiment.

(ii) For $t_1\ll V$ excitons with a dispersion given by Eq.(\ref{Eex})  
form and 
show up as strong absorption peak in the optical conductivity.
The excitons are due to an attractive potential between domain-wall 
pairs $\sim V/8l$,
whose prefactor reflects the fractional charge of the DW's, and
result as a consequence of the long-range (repulsive) Coulomb interaction
between electrons.
The appearance of excitonic states, which arise from the effective attraction
between the fractionally charged DW's, was to our knowledge
not noted before. In contrast in the frequently studied 
model with only nearest neighbor interactions  there is no bound state
because of the absence of attractive interactions between DW pairs.
However, in that case interchain interactions may provide an alternative
mechanism for confinement, as noted recently in work by   
Bhaseen and Tsvelik\cite{Bhaseen04}.

(iii) Edge-sharing chain compounds have the unusual property that the magnitude
of the second neighbor hopping matrix element $t_2$ is larger than the
nearest neighbor matrix element $t_1$. 
While $t_2$ hopping processes are frustrated in the classical WL state 
with alternating charge order, these processes surprisingly
contribute strongly to 
the exciton dispersion. They lift the degeneracy of the exciton state and the lower
branch, c.f. Eq.(\ref{Eexl}),
leads to an exciton instability at about $t_{2,c}\sim 0.16 V$.
The CDW state beyond  $t_{2,c}$ has a modulation period
twice as large as that of the WL. The charge modulation is weak in this CDW
state as infered from the calculation of the static charge structure factor.

(iv) Interestingly the optical conductivity at finite temperature reveals
an in-gap absorption which reflects the transitions between the soft exciton
near $q=\pi/2$ and the domain wall continuum. Thus this in-gap absorption
may provide a further identification of the WL state and also
allow for an 
independent experimental determination of the matrix element $t_2$.

(v) Moreover, we have calculated the temperature dependence of the DC-conductivity 
of the generalized Wigner lattice
from  the low-frequency absorption. The data for $\sigma(T)$
shows a crossover from activated 
behavior at low temperature to a basically temperature independent conductivity
at high temperatures. It turned out that  $\sigma(T)$ can be described over the full
temperature range by an activated behavior characterized by an activation energy
$E_a$ and a temperature dependent mobility $\sim T^{-\alpha}$.
This implies a strong decrease of the mobility of domain walls with increasing
temperature.

Finally, doped edge-sharing
 compounds provide a unique opportunity to study the competition 
between two entirely different states, the classical WL dictated by the long-range 
Coulomb interaction and the CDW of quantum mechanical origin, i.e., resulting
from a Fermi surface instability.
These materials highlight  
the importance of long-range Coulomb interaction in strongly correlated systems, 
--- and provide a one-dimensional test ground for the study of
charge stripe formation.
We have analysed here the charge dynamics and aspects of transport for a
generalized 1D Wigner lattice in its most simple realization, namely at
quarter-filling ($\rho=1/2$). Other commensurabilities are more complex and
show a hierachy of different charge excitations. Work along these lines
is in progress, as well as work 
on the effect of spin-charge coupling
particularly on the low-energy charge response.

\section*{Acknowledgements}
We like to thank  M. Jansen,  
G. Khaliullin, B. Keimer, W. Metzner,
A. Mishra and R. Zeyher for useful discussions.



\begin{thebibliography}{99}
\bibitem{modulation}
For electron concentration
$\rho=m/n$ the WL unit cell size is $n$ lattice units, as for a 
$4 k_F$ CDW with density modulation
$\rho(l)=\rho_0+\rho_1 cos(4 k_F l)$ and $4k_F=2\pi m/n$.
\bibitem{kf}
Instead of $k_F$ there are now two Fermi momenta $k_{F,1}$ and $k_{F,2}$
which depend on the electron concentration and on the ratio $t_2/t_1$.
\bibitem{melting}
As in the generalized Wigner lattice electrons are confined to one-dimension,
melting is not associated with the vanishing of the shear force but rather
with the disappearance of long-range correlations.
\bibitem{Tm}
We associate the melting temperature $T_m$ ad hoc, but consistent with
the study of Ref.\onlinecite{Erwin05}, with a 10 \% reduction of the
structure factor $N(q=\pi)$ compared to its zero temperature value.
\end{thebibliography}
\end{document}